\documentclass[conference]{IEEEtran}
\IEEEoverridecommandlockouts
\usepackage{cite}
\usepackage{amsmath,amssymb,amsfonts}
\usepackage{graphicx}
\usepackage{textcomp}
\usepackage{xcolor}
\usepackage{flushend}
\usepackage{balance}
\usepackage{epsfig}
\usepackage{epstopdf}

\usepackage[linesnumbered,ruled,vlined]{algorithm2e}
\usepackage{float}
\usepackage{subfig}
\usepackage{times}
\usepackage{url}

\usepackage{bm}
\usepackage{dsfont} 
\newcommand{\Id}{\bm{I}}
\newcommand{\one}{\mathds{1}}

\newcommand{\fDr}{\Delta f_{\mathrm{D}}}

\usepackage{tikz}
\usetikzlibrary{arrows.meta,positioning,fit,calc}

\def\BibTeX{{\rm B\kern-.05em{\sc i\kern-.025em b}\kern-.08em
    T\kern-.1667em\lower.7ex\hbox{E}\kern-.125emX}}

\begin{document}

\title{DRL-Based Beam Positioning for LEO Satellite Constellations with Weighted Least Squares 
\thanks{This work was supported in part by the National Science and Technology Council (NSTC) of Taiwan under Grant  113-2926-I-001-502-G and 114-2221-E-003-033.}
}
\author{\IEEEauthorblockN{Po-Heng Chou$^{1,3}$, Chiapin Wang$^{2}$, Kuan-Hao Chen$^{2}$, and Wei-Chen Hsiao$^{2}$}
\IEEEauthorblockA{
$^{1}$Research Center for Information Technology Innovation (CITI), Academia Sinica (AS), Taipei 11529, Taiwan\\
$^{2}$Department of Electrical Engineering, National Taiwan Normal University (NTNU), Taipei 106308 Taiwan\\
$^{3}$Bradley Department of Electrical and Computer Engineering (ECE), Virginia Tech (VT), Alexandria, VA 22305, USA\\
E-mails: d00942015@ntu.edu.tw, chiapin@ntnu.edu.tw, 61375063h@ntnu.edu.tw, 61275046h@ntnu.edu.tw\vspace{-0.2in}
}
}

\maketitle

\begin{abstract}
This paper investigates a lightweight deep reinforcement learning (DRL)-assisted weighting framework for CSI-free multi-satellite positioning in LEO constellations, where each visible satellite provides one serving beam (one pilot response) per epoch.
A discrete-action Deep Q-Network (DQN) learns satellite weights directly from received pilot measurements and geometric features, while an augmented weighted least squares (WLS) estimator provides physics-consistent localization and jointly estimates the receiver clock bias.
The proposed hybrid design targets an accuracy-runtime trade-off rather than absolute supervised optimality. 
In a representative 2-D setting with 10 visible satellites, the proposed approach achieves sub-meter accuracy (0.395m RMSE) with low computational overhead, supporting practical deployment for resource-constrained LEO payloads.
\end{abstract}


\section{Introduction}
The integration of terrestrial, aerial, and satellite segments into a unified ground-air-space architecture has emerged as a key enabler for future sixth-generation (6G) networks, promising seamless connectivity, low latency, and global coverage~\cite{Saleh2025_TNNTN}. Among these, low Earth orbit (LEO) satellite constellations are particularly attractive due to their wide coverage, rapid revisit capability, and suitability for delay-sensitive services. However, their highly dynamic topology, interference-prone links, and constrained on-board processing capabilities present significant challenges for robust and energy-efficient service delivery~\cite{Zhu2024_embedding, Saleh2025_TNNTN, Cao2025_symbiotic}.

Recent research has addressed these challenges from multiple perspectives. A Lyapunov-based beam management strategy was developed for LEO networks with random traffic arrivals and time-varying topologies, effectively reducing beam revisit times and handover frequency~\cite{Zhu2024_embedding}. A multi-satellite beam-hopping framework was introduced to balance load and avoid interference in non-geostationary orbit (NGSO) constellations, significantly improving spectral efficiency~\cite{Lin2023_multipath}. Deep reinforcement learning (DRL)-driven handover protocols have been designed to eliminate measurement-report overhead, thereby reducing access delay and collision rates in regenerative-type LEO networks~\cite{Lee2023_handovers}. Cooperative beam scheduling and beamforming approaches in multi-beam LEO systems have shown positioning accuracy gains of up to 17.1\% under Cramér-Rao lower bound (CRLB) criteria~\cite{xv2024_tvt}. Reinforcement learning-based robust beamforming methods for multibeam downlinks have also demonstrated resilience to imperfect channel knowledge~\cite{Schroder2024_flexible}. In addition, cooperative beam-hopping control strategies in ultra-dense LEO constellations have been proposed to enhance time-difference-of-arrival (TDOA) positioning performance~\cite{Wang2021_coop}.

Beyond the physical layer, system-level frameworks have also been explored. A symbiotic radio approach using collaborative DRL has been proposed for intelligent resource optimization in non-terrestrial networks (NTNs)~\cite{Cao2025_symbiotic}. Standardization and system-level research on integrated terrestrial and non-terrestrial networks (TN-NTN) localization have highlighted key challenges in synchronization, interference, and robustness~\cite{Saleh2025_TNNTN}. Moreover, federated learning frameworks have been investigated for NTNs to enhance scalability and privacy in multi-tier deployments~\cite{Farajzadeh2025_FLNTN}.

Despite these advances, most existing solutions still rely heavily on explicit channel state information (CSI) or involve computationally intensive optimization, limiting real-time applicability for resource-constrained LEO platforms. Traditional statistical estimation methods, such as weighted least squares (WLS)~\cite{Kay1993_estimation}, widely used in positioning problems, provide reliable accuracy but lack adaptability in dynamic multi-satellite environments with heterogeneous link qualities. Meanwhile, DRL has proven effective in sequential decision-making tasks~\cite{Mnih2015_DQN}, yet its integration with lightweight positioning remains underexplored in NTNs.

Motivated by these gaps, this work extends the use of DRL beyond prior studies that focused primarily on handover optimization~\cite{Lee2023_handovers}, load-balanced beam hopping~\cite{Lin2023_multipath}, or cooperative scheduling~\cite{xv2024_tvt}. We adopt the DQN framework~\cite{Mnih2015_DQN}, which approximates the action-value function with deep neural networks to enable efficient decision-making in large state-action spaces. Unlike previous DRL approaches requiring explicit CSI or heavy network coordination, the proposed framework applies DQN for multi-satellite positioning in multiple-input multiple-output (MIMO) LEO systems and integrates it with a WLS estimator~\cite{Kay1993_estimation}. This design allows adaptive beam selection and weighting directly from received pilot measurements. Thus, it eliminates explicit CSI estimation, achieving robustness under interference-prone and global navigation satellite system (GNSS)-denied conditions. In addition, it significantly reduces computational complexity compared to conventional optimization or deep learning baselines.
While CRLB-based optimization offers theoretical bounds, it is analytically intractable for fast-varying satellite geometry. Hence, an augmented WLS estimator is adopted, whose efficiency asymptotically approaches the CRLB under Gaussian noise.

The main contributions of this paper are summarized as follows:
\begin{itemize}
    \item \textbf{CSI-free, lightweight hybrid design:} We couple a discrete-action DQN \emph{satellite-weighting} policy with an augmented WLS solver to enable CSI-free multi-satellite positioning, while maintaining a small on-board inference footprint.
    \item \textbf{Deployment-oriented action parameterization:} We design a quantized \emph{satellite-weight} action space compatible with practical on-board control granularity, improving numerical robustness of the downstream WLS inversion while remaining tractable for DQN training.
    \item \textbf{Accuracy-runtime evidence and insight:} We provide a comparative evaluation highlighting the trade-off between positioning accuracy and computational cost across representative baselines, illustrating the regime where the proposed approach is attractive for real-time LEO operation.
\end{itemize}

\section{System Model}

\begin{figure}[t]
    \centering
    \includegraphics[width=\linewidth]{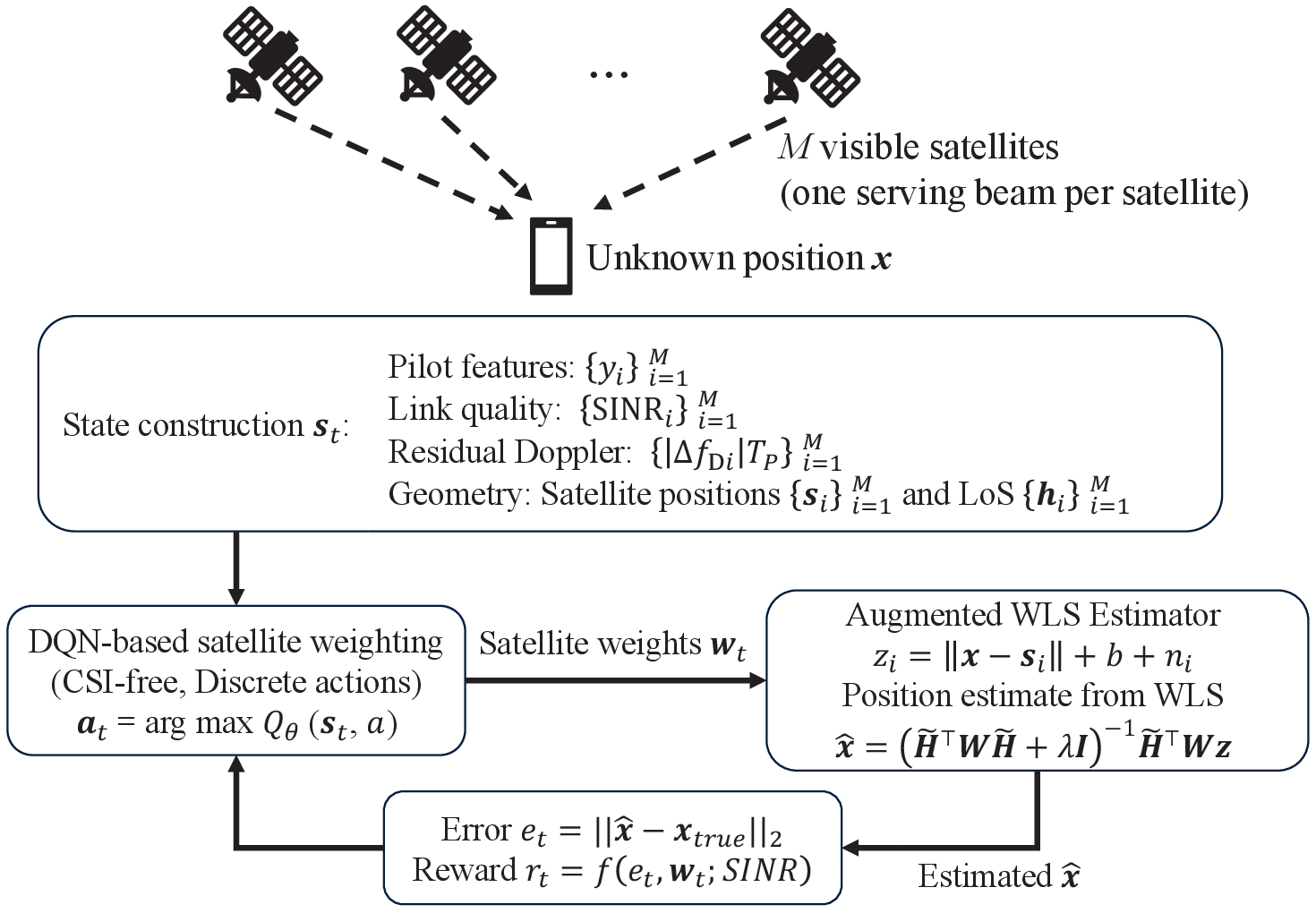}
    \caption{System overview of the proposed CSI-free DQN-WLS multi-satellite positioning framework.}
    \label{fig:system_overview}
    \vspace{-0.1in}
\end{figure}

Fig.~\ref{fig:system_overview} illustrates the proposed CSI-free multi-satellite positioning pipeline.
At each epoch, $M$ visible LEO satellites are available, and each satellite schedules one serving beam to deliver a pilot observation.
Pilot-derived features $\{y_i\}_{i=1}^{M}$, geometry information (e.g., $\{\bm{s}_i,\bm{h}_i\}$), and the normalized residual Doppler magnitudes $\{|{\fDr}_i|T_p\}_{i=1}^{M}$ (with $T_p$ defined in~\eqref{eq:sigma_doppler}) constitute the DQN state.
The DQN selects a discrete action that is mapped to a normalized satellite-weight vector, which is then applied to the augmented WLS estimator to jointly estimate the UT position and the receiver clock bias.

\subsection{Coordinate System and Satellite geometry}
We consider $M$ visible LEO satellites from a constellation.
Let $\{\bm{s}_i\}_{i=1}^{M}$ denote the satellite positions in the earth-centered, earth-fixed (ECEF) coordinate system within a short observation window.
The unknown UT position is denoted by $\bm{x}\in\mathbb{R}^{d\times 1}$.

For satellite $i$, the line-of-sight (LoS) unit vector from the satellite to the UT is
\begin{equation}
\bm{h}_i \triangleq \frac{\bm{x}-\bm{s}_i}{\|\bm{x}-\bm{s}_i\|}, \quad i=1,\ldots,M.
\label{eq:los}
\end{equation}
where $\bm{h}_i \in \mathbb{R}^{d\times1}$ and $d=2$ or $3$ depending on the positioning dimension.
The satellite geometry $\{\bm{s}_i,\bm{h}_i\}$ is assumed quasi-static within a 1 to 5 s window, updated by ephemeris and the scheduler.

\subsection{Channel Model}
Following the model in~\cite{xv2024_tvt}, the narrowband line-of-sight (LoS) channel between the satellite and the UT is dominated by a single deterministic path.
Let $\bm{s}_i\in\mathbb{R}^3$ denote the satellite antenna phase-center position and $d_i=\|\bm{x}-\bm{s}_i\|$ the slant range between satellite $i$ and the UT.
Consider a uniform planar array (UPA)~\cite{Shafin2017_3DMIMO} with $N_x\times N_y$ elements and total antenna count $N=N_xN_y$.
The complex baseband channel vector from satellite $i$ is expressed as
\begin{equation}
    \bm{g}_i = \beta_i e^{j\phi_i} \bm{a}_t(\varphi_i) \in \mathbb{C}^{N \times 1},
    \label{eq:channel_vector}
\end{equation}
where $\phi_i$ denotes the random phase uniformly distributed in $[0,2\pi)$, 
$\bm{a}_t(\varphi_i)$ is the UPA transmit steering vector at the departure angle $\varphi_i$, 
and $\beta_i$ represents the large-scale channel gain defined as
\begin{equation}
    \beta_i = \sqrt{\frac{G_t G_r}{L_i}}, 
    \qquad 
    L_i = \left(\frac{4\pi f_c d_i}{c}\right)^{2},
    \label{eq:beta_gain}
\end{equation}
with $f_c$ denoting the carrier frequency, $c$ the speed of light, 
and $G_t, G_r$ the transmit and receive antenna gains, respectively.
The transmit steering vector $\bm{a}_t(\varphi_i)$ follows a standard $N_x\times N_y$ UPA response model \cite{Shafin2017_3DMIMO}. 
To reduce exposition, we omit its element-wise expression and directly use $\bm{a}_t(\varphi_i)$ to compute the effective beamformed gain $|\bm{g}_i^{H}\bm{f}_i|^2$ in \eqref{eq:sinr}.

Within each epoch, the UT correlates $N_p$ received pilot samples to obtain one scalar pilot feature per satellite:
\begin{equation}
y_i \triangleq \frac{1}{N_p}\sum_{n=0}^{N_p-1} r[n]\,x_i^{*},
\label{eq:pilot_feature}
\end{equation}
where $r[n]$ denotes the received pilot sample and $x_i$ is the known pilot symbol of satellite $i$. 
After ephemeris-based Doppler pre-compensation, the residual Doppler is modeled as ${\fDr}_i\sim\mathcal{N}(0,\sigma_{\fDr}^2)$, i.i.d.\ across satellites and epochs.

Accordingly, we compute the instantaneous power-domain link-quality metric under frequency reuse as
\begin{equation}
    \mathrm{SINR}_i = 
    \frac{|\bm{g}_i^{\!H}\bm{f}_i|^2}
    {\sum_{k\neq i} |\bm{g}_k^{\!H}\bm{f}_k|^2 + \sigma^2},
    \label{eq:sinr}
\end{equation}
where the summation term accounts for \emph{frequency-reuse co-channel beams} transmitted by other satellites on the same time-frequency resource, and $\bm{g}_k$ denotes the channel vector from satellite $k$ to the UT within the considered epoch.
The residual Doppler impact is incorporated via an equivalent measurement-reliability penalty in \eqref{eq:sinr_eff}-\eqref{eq:sigma_doppler}. 
We emphasize that a full orthogonal frequency-division multiplexing (OFDM) inter-carrier interference (ICI) treatment is beyond the scope of this work and can be captured by an equivalent SINR/variance penalty when needed.
The instantaneous per-satellite link-quality metric $\mathrm{SINR}_i$ under frequency reuse computed from~\eqref{eq:channel_vector}-\eqref{eq:sinr} 
is normalized to $[0,1]$ and later used as part of the DRL observation vector.

\subsection{Observation Model}
In conventional multi-satellite positioning, the UT location is inferred from multiple range-like measurements collected from visible satellites.
To build a unified and analytically tractable model, we define
a set of differentiable measurements $\bm{z}\in\mathbb{R}^{M}$ that depend on the geometric distances between $\bm{x}$ and the satellite positions $\{\bm{s}_i\}$.
For satellite $i$, a pseudorange-like measurement is modeled as
\begin{equation}
z_i = \|\bm{x}-\bm{s}_i\| + b + n_i,
\label{eq:pseudo}
\end{equation}
where $b$ is the receiver clock bias and $n_i\sim\mathcal{N}(0,\sigma_i^2)$ is zero-mean Gaussian noise whose variance reflects the effective link quality of satellite~$i$.
To incorporate residual Doppler, we model an equivalent reliability penalty on the range-like measurement. Specifically, we define an \emph{effective} SINR as
\begin{equation}
\mathrm{SINR}_i^{\mathrm{eff}}
=
\frac{\mathrm{SINR}_i}
{1+\kappa\left(|{\fDr}_i|T_p\right)^2},
\label{eq:sinr_eff}
\end{equation}
where $\kappa$ is a scaling parameter controlling the Doppler-induced reliability penalty.
The measurement noise variance is modeled as
\begin{equation}
\sigma_i^2
=
\frac{\sigma_0^2}{\mathrm{SINR}_i^{\mathrm{eff}}},
\qquad
T_p=N_p T_s,
\label{eq:sigma_doppler}
\end{equation}
where $\sigma_0^2$ is a nominal noise level, $\kappa$ controls the Doppler-induced penalty strength, and $T_p$ denotes the pilot integration time used for pilot correlation/averaging within one epoch. This model captures that residual Doppler degrades pilot correlation quality and thus reduces the reliability of range-like measurements, which is directly reflected in the WLS weighting.

Linearizing the pseudorange-like measurements in \eqref{eq:pseudo} at a reference $\bm{x}_0$ yields
\begin{equation}
    \bm{z} \approx \bm{H}\bm{x} + b\one + \bm{n},
    \label{eq:linear}
\end{equation}
where the $i$-th row of $\bm{H}\in\mathbb{R}^{M\times d}$ is $\bm{h}_i^\top$ defined in \eqref{eq:los}, $\one\in\mathbb{R}^{M\times 1}$, and $\bm{x}_0$ is an initial position estimate.

\subsection{Augmented Weighted Least Squares (WLS)}
Let $\tilde{\bm{H}}=[\bm{H}\ \ \one]\in\mathbb{R}^{M\times(d+1)}$ and $\tilde{\bm{x}}=[\bm{x}^\top\ b]^\top$.
Given $\bm{W}=\mathrm{diag}(w_1,\ldots,w_M)$, the augmented WLS estimate is
\begin{equation}
\hat{\tilde{\bm{x}}}
=
\big(\tilde{\bm{H}}^{\top}\bm{W}\tilde{\bm{H}}+\lambda \Id\big)^{-1}
\tilde{\bm{H}}^{\top}\bm{W}\bm{z},
\label{eq:awls}
\end{equation}
with ridge parameter $\lambda>0$ for numerical stability. We then parse $\hat{\bm{x}}$ and $\hat{b}$ from $\hat{\tilde{\bm{x}}}$.

\section{Problem Formulation}

From the system model above, the WLS estimator in~\eqref{eq:awls}
computes the UT position $\hat{\bm{x}}$ based on the satellite weight vector 
$\bm{w}=[w_1,\ldots,w_M]^{\top}$.
The instantaneous positioning error is defined as
\begin{equation}
    e = \|\hat{\bm{x}}-\bm{x}_{\text{true}}\|_2.
    \label{eq:error}
\end{equation}
The objective of the multi-satellite positioning problem is to find the optimal weight vector $\bm{w}^*$ that minimizes the expected positioning error:
\begin{equation}
    \bm{w}^{*}
    = \arg\min_{\bm{w}\in\mathcal{W}} 
    \mathbb{E}\!\left[\|\hat{\bm{x}}(\bm{w})-\bm{x}_{\text{true}}\|_2^2\right],
    \label{eq:opt_problem}
\end{equation}
subject to $w_i\in[0,1]$ and $\sum_i w_i=1$,
where $\mathcal{W}$ denotes the feasible set of normalized weight vectors.

The above optimization is nonlinear and highly dependent on the satellite geometry through the matrix $\bm{H}$ in~\eqref{eq:linear}.
Since $\bm{H}$ varies rapidly with satellite motion and the time-varying set of visible/serving satellite links, solving~\eqref{eq:opt_problem} analytically or by exhaustive search is computationally prohibitive.
Conventional heuristic methods assign fixed or power-based weights 
that cannot adapt to dynamic link conditions.

To overcome these limitations, we reformulate the satellite-weight optimization as a sequential decision-making problem, where an agent learns to adjust $\bm{w}$ based on observed satellite-link features and positioning feedback.
This motivates the proposed DRL approach detailed in Section~\ref{sec:proposed_dqn}.
To support practical deployability and numerical stability, the continuous weight vector is quantized into a finite action set: (i) it matches hardware-friendly weight granularity (e.g., codebook-like control), (ii) it mitigates ill-conditioning in the augmented WLS inversion under noisy beams, and (iii) it yields a tractable discrete action space for DQN training.

\section{DQN-WLS Based Multi-Satellite Positioning}
\label{sec:proposed_dqn}

In this section, we present a DRL framework that integrates a DQN with a WLS estimator for accurate and adaptive multi-satellite positioning in LEO satellite systems. 
Unlike conventional geometry- or CSI-based methods, the proposed DQN-WLS scheme learns satellite-weighting strategies directly from received pilot measurements and geometric features, thus eliminating explicit CSI estimation and reducing computational complexity.

\subsection{Markov Decision Process Formulation}
The beam-weight optimization problem in~\eqref{eq:opt_problem} is reformulated as a Markov decision process (MDP) $\langle\mathcal{S},\mathcal{A},\mathcal{P},\mathcal{R},\gamma\rangle$.  
At each discrete time step $t$, the environment (\textit{BeamEnv}) generates a state vector $\bm{s}_t\in\mathcal{S}$ that encapsulates both geometric and signal-level features of $M$ visible satellite links.  
These include normalized satellite distances (or slant ranges), azimuth-elevation indicators, per-satellite SINR metrics, previous residual errors, and prior weight distributions.

\noindent\textbf{Action parameterization.}
To comply with DQN's discrete-action setting, we quantize the continuous satellite weights into a finite action set constructed by top-$K$ satellite selections and coarse weight levels $\{0,0.25,0.5,0.75,1\}$, followed by $\ell_1$ normalization. Near-duplicate actions are pruned by correlation screening to keep the action set tractable.

The agent selects a discrete action $a_t\in\mathcal{A}$, which is mapped to a normalized weight vector $\bm{w}_t$ applied to the WLS estimator.
After performing the WLS estimation $\hat{\bm{x}}_t$, the environment computes the positioning error $e_t=\|\hat{\bm{x}}_t-\bm{x}_{\text{true}}\|_2$ and generates a scalar reward $r_t$ that reflects both accuracy and stability:
\begin{equation}
r_t = -\left(\frac{e_t}{\tau}\right)^{2}
      - \alpha \sum_{i=1}^{M} w_{t,i}\big(1-\mathrm{SINR}_i^{\mathrm{eff}}\big)
      - \beta \big(1-\mathrm{Entropy}(\bm{w}_t)\big),
\label{eq:reward}
\end{equation}
where the entropy term is defined as
\begin{equation}
\mathrm{Entropy}(\bm{w}_t) = -\sum_{i=1}^{M} w_{t,i} \log w_{t,i},
\end{equation}
which encourages exploration by promoting weight diversity. 
Here, $\tau$ controls the error-scaling factor, $\alpha$ penalizes low-quality satellite links, and $\beta$ regulates the entropy contribution.
The discount factor $\gamma=0.99$ encourages long-term optimization over sequential interactions.

\subsection{Network Architecture}
The DQN approximates the action-value function $Q_{\theta}(\bm{s},\bm{a})$ parameterized by $\theta$. 
The input layer concatenates per-satellite features with a total feature dimension $F=60$ for $M=10$ satellites.
Two fully connected hidden layers of 128 neurons each, activated by ReLU, are used to extract latent geometric representations. 
After pruning redundant combinations, the discrete action set contains approximately $10^{2}$ valid satellite-weight configurations. 
The network outputs $|\mathcal{A}|$ Q-values over these quantized actions, and the selected action is mapped to a normalized weight vector $\bm{w}_t$.

To stabilize the subsequent WLS inversion, a ridge term $\lambda \bm{I}$ is added to~\eqref{eq:awls}.  
The network parameters are initialized using He initialization, optimized via Adam with learning rate $\eta=10^{-3}$, and updated using the Huber loss.  
A target network $Q_{\bar{\theta}}$ is periodically synchronized with the main network $Q_{\theta}$ every $C=200$ iterations to mitigate oscillations in temporal-difference targets.

\subsection{Training Procedure}
During training, the agent interacts with the environment over multiple episodes.  
At each step $t$, the agent observes the current state $\bm{s}_t$, selects an action $\bm{w}_t=f_{\theta}(\bm{s}_t)$ according to an $\epsilon$-greedy policy, applies $\bm{w}_t$ to the WLS estimator, and receives a reward $r_t$ and next state $\bm{s}_{t+1}$.  
Each transition $(\bm{s}_t,\bm{w}_t,r_t,\bm{s}_{t+1})$ is stored in a replay buffer $\mathcal{D}$ of size $10^4$.  
Mini-batches of 32 samples are randomly drawn from $\mathcal{D}$ to update the network parameters.  

The target value for each sample is defined as
\begin{equation}
    y_t = r_t + \gamma\, Q_{\bar{\theta}}\!\big(\bm{s}_{t+1}, f_{\bar{\theta}}(\bm{s}_{t+1})\big),
    \label{eq:td_target}
\end{equation}
and the temporal-difference (TD) loss follows the Huber formulation:
\begin{equation}
L_t(\theta) =
\begin{cases}
\frac{1}{2}\delta_t^2, & |\delta_t|<1,\\
|\delta_t|-\frac{1}{2}, & \text{otherwise,}
\end{cases}
\label{eq:huber_loss}
\end{equation}
where $\delta_t = y_t - Q_{\theta}(\bm{s}_t,\bm{w}_t)$.  
Gradients are clipped with $\|\nabla_{\theta}\|\le5$ to prevent divergence, and $\epsilon$ decays exponentially from 1.0 to 0.01 during training.  
Convergence is typically achieved within 800 to 1000 episodes in the considered settings, after which the positioning error enters a stable sub-meter regime.
Furthermore, the learned satellite weights exhibit interpretable behavior: high-SINR beams tend to receive larger weights, while redundant or correlated beams are automatically suppressed, indicating that the agent learns beam reliability from pilot-level observations without explicit CSI modeling.

\begin{algorithm}[t]
\caption{DQN-WLS Based Satellite Positioning}
\SetAlgoLined
\KwIn{Environment BeamEnv, learning rate $\eta$, discount factor $\gamma$, target update interval $C$, exploration schedule $\epsilon$}
\KwOut{Trained DQN policy $\pi_{\theta}$ for adaptive beam-weighting}
Initialize replay buffer $\mathcal{D}$, networks $Q_{\theta}$ and target $Q_{\bar{\theta}} \!\leftarrow\! Q_{\theta}$\;
\For{episode $=1$ to $E_{\max}$}{
    Reset environment and obtain initial state $s_0$\;
    \For{each time step $t$}{
        Select action $\bm{w}_t \!\leftarrow\! f_{\theta}(s_t)$ and apply $\epsilon$-greedy exploration\;
        Normalize $\bm{w}_t$ s.t. $\sum_i w_{t,i}=1$, form $W_t=\mathrm{diag}(\bm{w}_t)$\;
        Compute $\hat{\bm{x}}_t$ via WLS in~\eqref{eq:awls}\;
        Compute positioning error $e_t$ using~\eqref{eq:error}\;
        Evaluate reward $r_t$ via~\eqref{eq:reward} and observe next state $s_{t+1}$\;
        Store $(s_t,\bm{w}_t,r_t,s_{t+1})$ in buffer $\mathcal{D}$\;
        Sample mini-batch from $\mathcal{D}$\;
        Compute target $y_t$ using~\eqref{eq:td_target}\;
        Update $\theta \!\leftarrow\! \theta - \eta \nabla_{\theta} L_t(\theta)$ via Huber loss~\eqref{eq:huber_loss}\;
        \If{mod($t,C$)$=0$}{update target network $Q_{\bar{\theta}} \!\leftarrow\! Q_{\theta}$\;}
        \If{convergence or terminal state reached}{break\;}
    }
}
\Return{Optimal policy $\pi_{\theta}^*$ generating $\bm{w}_t^*$ for each $s_t$}
\end{algorithm}

\section{Simulation Results and Analysis}

\subsection{Experimental Setup}
All simulations were conducted in Python using the PyTorch 2.5 framework on a workstation equipped with an Intel i9 CPU and an NVIDIA RTX A6000 GPU. 
A 2-D coverage area of $1000\times1000$~m$^2$ is considered with $M=10$ visible satellites. 
The discount factor is set to $\gamma=0.99$, the learning rate $\eta=10^{-3}$ (Adam optimizer), and the target network update interval $C=200$. 
Each training episode consists of $T=100$ steps, and the replay buffer stores $10^4$ transitions.

\noindent\textbf{Residual Doppler.}
After ephemeris-based Doppler pre-compensation, the residual Doppler is modeled as a zero-mean Gaussian random variable with variance $\sigma_{\fDr}^{2}=5$ (Hz$^2$), and its impact is incorporated into the measurement reliability via \eqref{eq:sinr_eff}--\eqref{eq:sigma_doppler}, where $T_p=N_pT_s$.

We compare six methods: PPO \cite{Schulman2017_PPO}, DDQN \cite{VanHasselt2016_DDQN}, DQN \cite{Mnih2015_DQN}, ALG-B \cite{Elsanhoury2024_BeamBased}, LSTM-WLS \cite{Sbeity2023_LSTM}, and the proposed DQN-WLS.
The implementation of all six algorithms, including network configurations and simulation parameters, is available for reproducibility.\footnote{Source code and configuration files are available at: \texttt{github.com/BoneZhou/DRL-LEO-Beam-Positioning}}

\subsection{Convergence Behavior}
\begin{figure}[t]
    \centering
    \subfloat[PPO~\cite{Schulman2017_PPO}]{\includegraphics[width=0.45\textwidth]{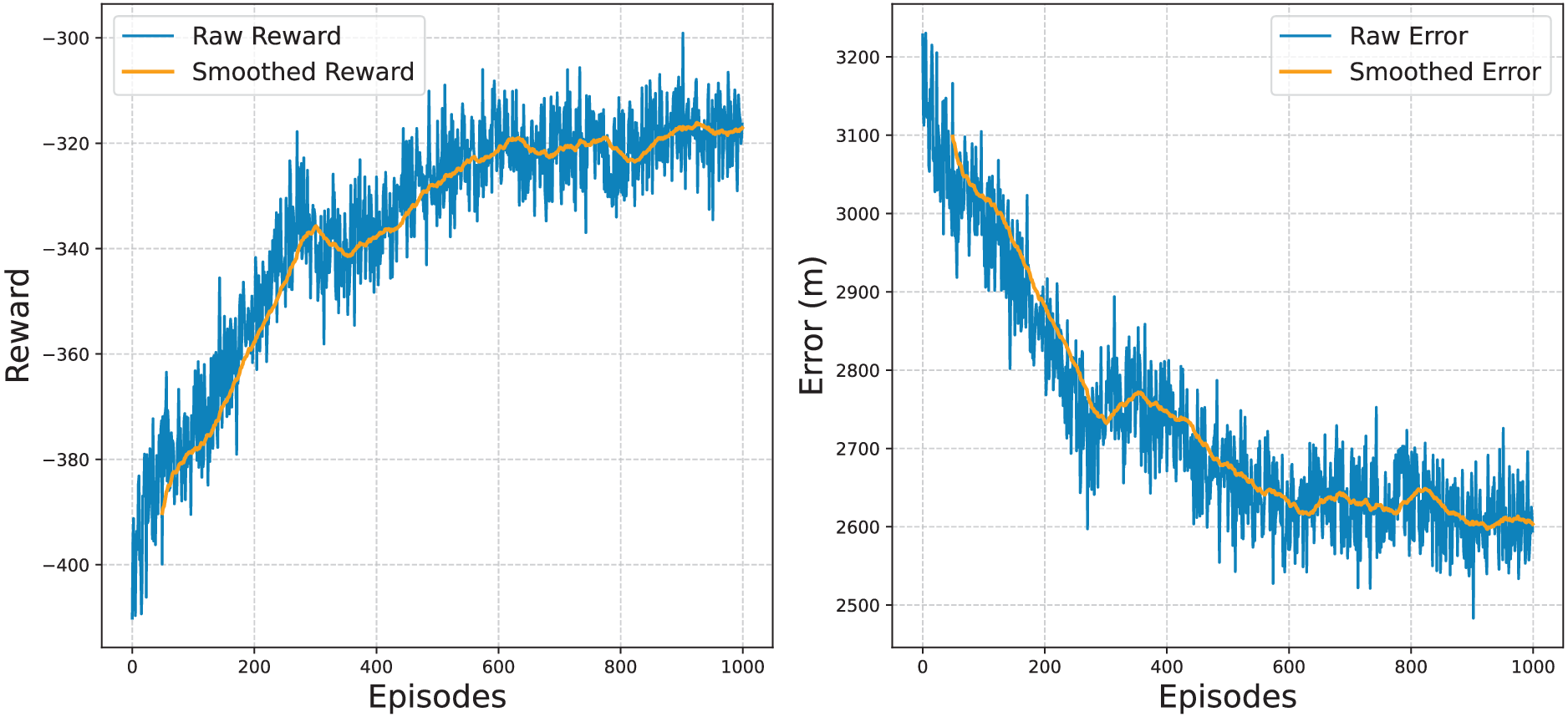}}\\
    \subfloat[DDQN~\cite{VanHasselt2016_DDQN}]{\includegraphics[width=0.45\textwidth]{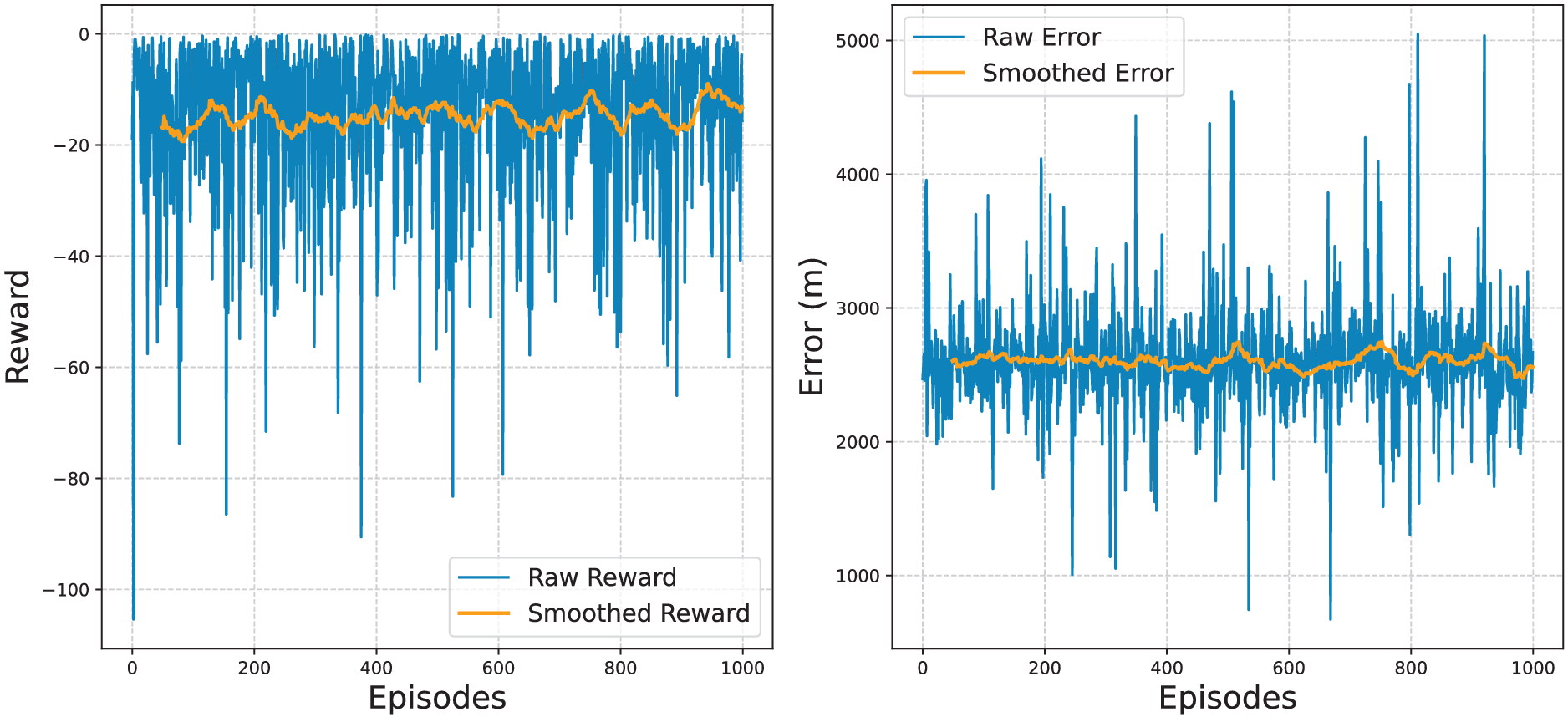}}\\
    \subfloat[DQN~\cite{Mnih2015_DQN}]{\includegraphics[width=0.45\textwidth]{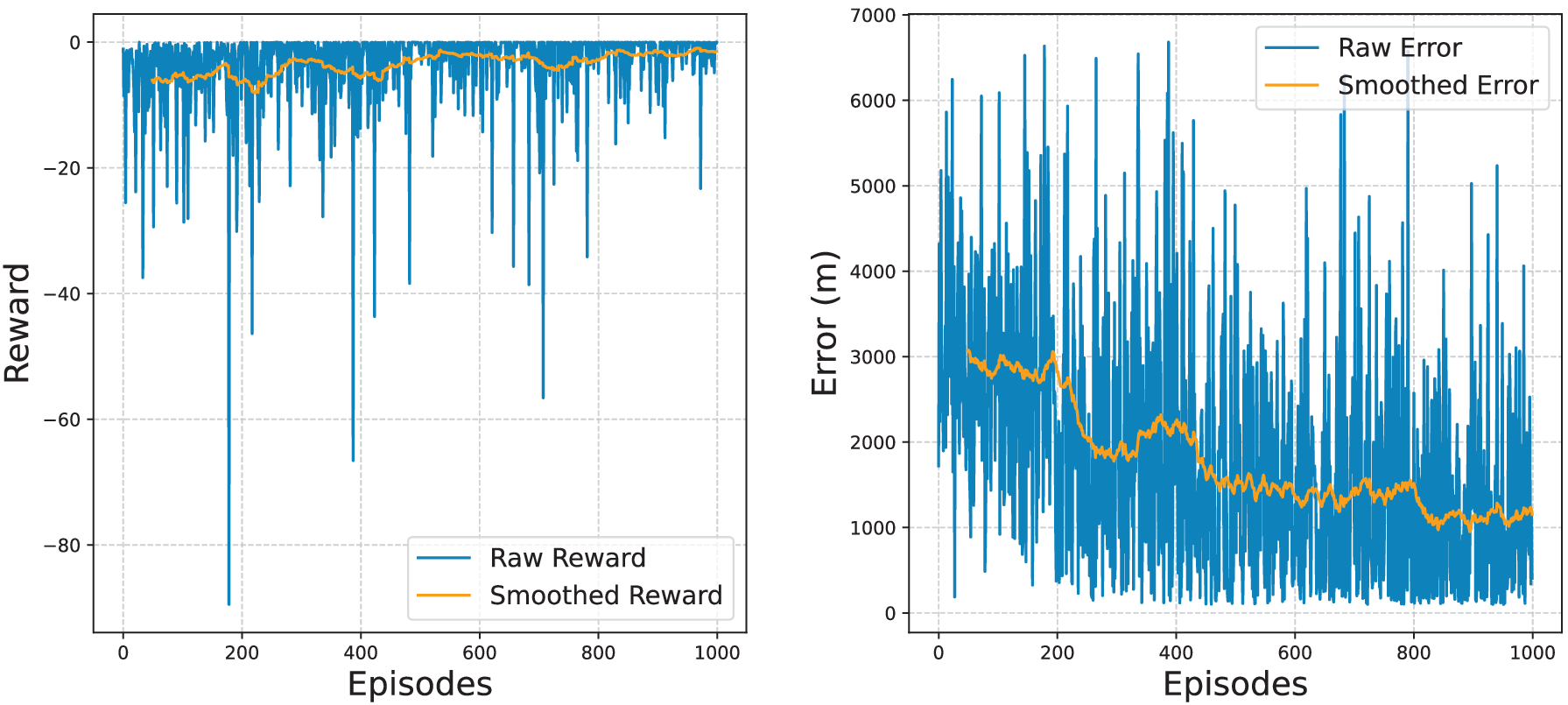}}\\
    \subfloat[Proposed DQN-WLS]{\includegraphics[width=0.45\textwidth]{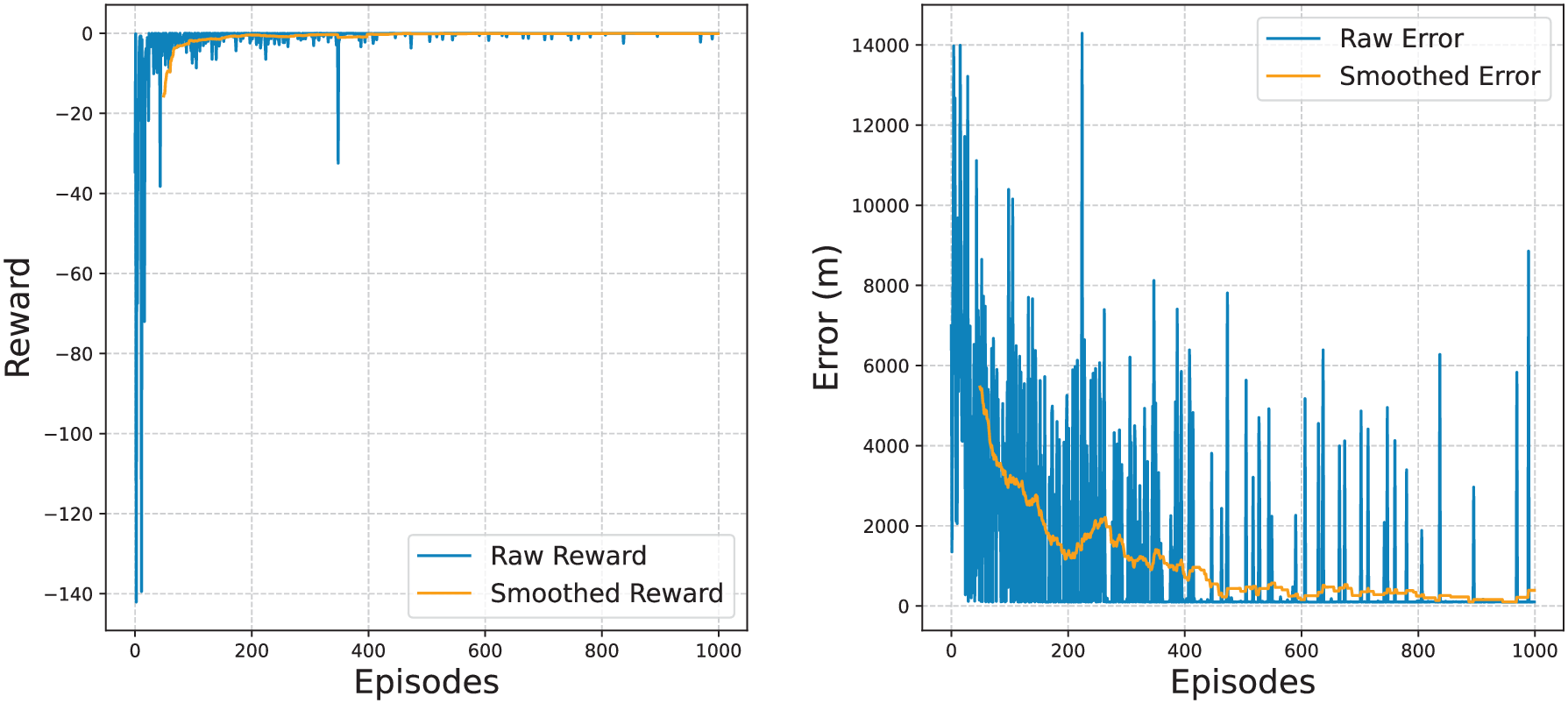}}
    \caption{Training convergence of cumulative reward and positioning error for PPO, DDQN, DQN, and the proposed DQN-WLS.}
    \label{fig:conv_all}
\vspace{-0.3in}
\end{figure}

Fig.~\ref{fig:conv_all} reports the learning curves of PPO, DDQN, DQN, and the proposed DQN-WLS.
The proposed DQN-WLS converges stably and achieves the lowest positioning error among the evaluated DRL baselines, benefiting from the geometry-consistent WLS refinement.

\subsection{Positioning Visualization}
\begin{figure*}[t]
    \centering
    \subfloat[ALG-B~\cite{Elsanhoury2024_BeamBased}]{\includegraphics[width=0.22\textwidth]{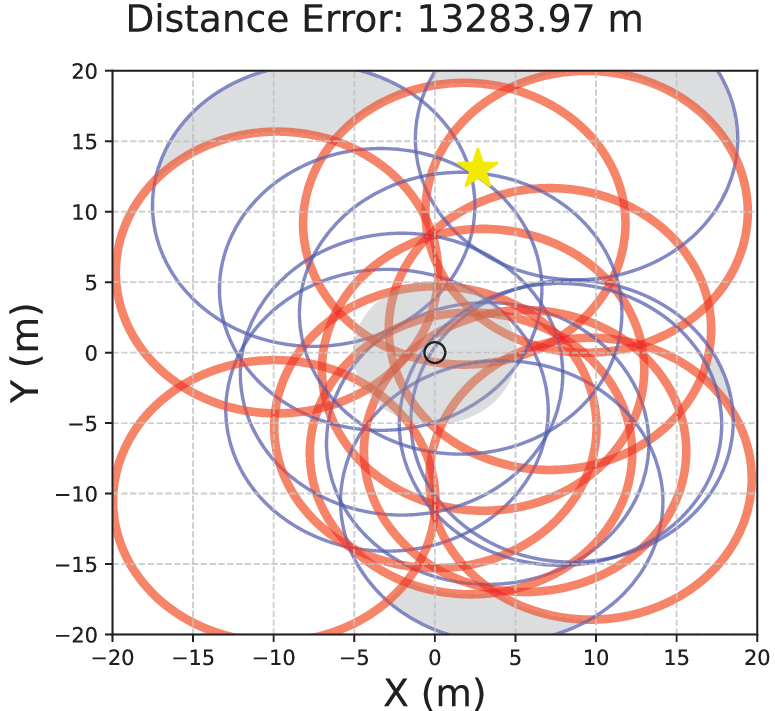}}
    \hfil
    \subfloat[DQN~\cite{Mnih2015_DQN}]{\includegraphics[width=0.22\textwidth]{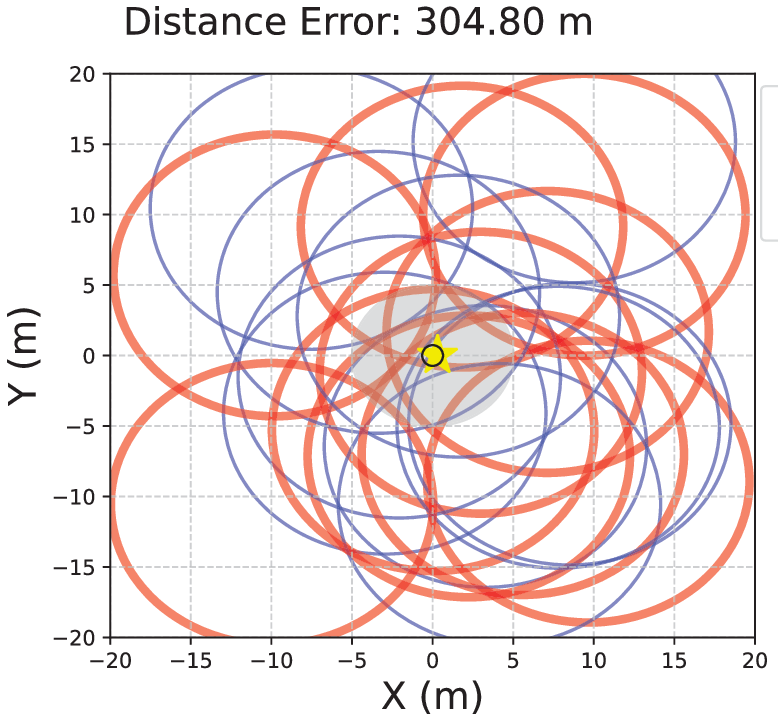}}
    \hfil
    \subfloat[Proposed DQN-WLS]{\includegraphics[width=0.32\textwidth]{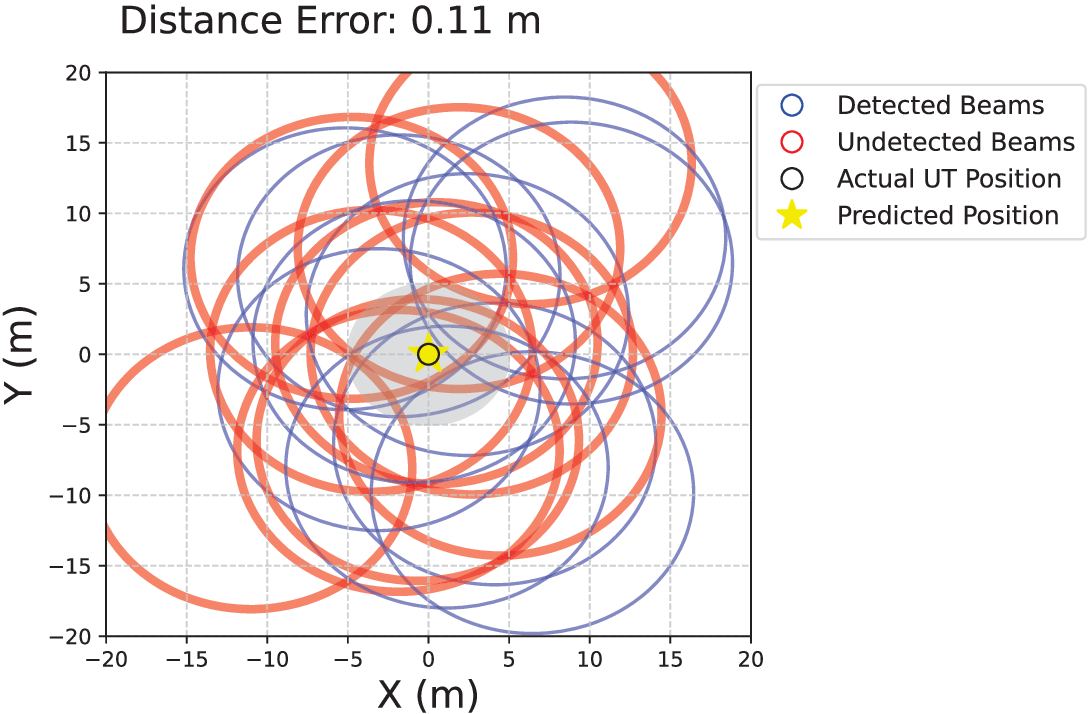}}
    \caption{Positioning visualization at Episode~1000 under identical satellite geometry and noise conditions. 
    The yellow star denotes the estimated UT position, and the black circle indicates the ground truth. 
    Compared with the geometry-driven ALG-B and standard DQN, the proposed DQN-WLS aligns closely with the ground truth and yields the smallest residual error.}
    \label{fig:pos_tri}
    \vspace{-0.2in}
\end{figure*}

Fig.~\ref{fig:pos_tri} compares ALG-B, DQN, and the proposed DQN-WLS at Episode~1000.
The proposed DQN-WLS closely matches the ground truth and achieves sub-meter accuracy (0.395~m RMSE), while ALG-B and DQN exhibit much larger deviations.

\subsection{Performance Comparison}
Table~\ref{tab:model_comp} summarizes the overall computational and localization performance of six representative algorithms, including both geometric and learning-based approaches.
The conventional beam-intersection method (ALG-B~\cite{Elsanhoury2024_BeamBased}) exhibits the largest localization deviation ($>3$~km) due to its reliance on fixed beam footprints without adaptive weighting.
Among DRL-based schemes, PPO~\cite{Schulman2017_PPO} and DDQN~\cite{VanHasselt2016_DDQN} incur substantially higher runtime under the considered training configuration, while the standard DQN~\cite{Mnih2015_DQN} improves efficiency and reduces the positioning error to 10.38~m.
The LSTM-WLS~\cite{Sbeity2023_LSTM} framework achieves extremely high accuracy (0.01~m) but incurs high training cost and longer inference latency, making it impractical for real-time onboard operation.
By contrast, the proposed DQN-WLS achieves sub-meter accuracy (0.395~m) with the lowest runtime (9.43~s per episode) and nearly identical computational complexity to DQN ($\approx$19~GFLOPs).
These results confirm that the proposed DQN-WLS achieves a favorable accuracy--efficiency trade-off with low per-step complexity.
In addition, we include an ablation where ${\fDr}_i=0$ (Doppler-free) and observe only a negligible RMSE change in our considered setting, indicating that the proposed method is robust under typical residual Doppler after compensation.

\begin{table}[t]
\caption{Comparison of Positioning and Learning-Based Methods}
\label{tab:model_comp}
\renewcommand{\arraystretch}{1}
\setlength{\tabcolsep}{1pt}
\begin{tabular}{lcccc}
\hline
\textbf{Method} & \textbf{FLOPs (M)} & \textbf{Total FLOPs (G)} & \textbf{Time (s)} & \textbf{RMSE (m)} \\
\hline
\textbf{ALG-B}~\cite{Elsanhoury2024_BeamBased} & -- & -- & 2.61 & 3688.14 \\
\textbf{PPO}~\cite{Schulman2017_PPO} & 0.07 & 92.87 & 3016.02 & 2558.33 \\
\textbf{DDQN}~\cite{VanHasselt2016_DDQN} & 0.04 & 77.59 & 350.08 & 2504.05 \\
\textbf{DQN}~\cite{Mnih2015_DQN} & 0.02 & 18.94 & 13.52 & 10.38 \\
\textbf{LSTM-WLS}~\cite{Sbeity2023_LSTM} & 31.24 & 78.09 & 262.96 & \textbf{0.01} \\
\textbf{Proposed DQN-WLS} & 0.02 & 18.95 & \textbf{9.43} & \textbf{0.395} \\
\hline
\end{tabular}
\end{table}

\noindent\textit{Note:} Time and RMSE are averaged over 10 seeds under identical settings.

\subsection{Discussion}
The favorable accuracy-runtime trade-off of DQN-WLS stems from its hybrid structure, combining the adaptivity of reinforcement learning with the stability and interpretability of model-based estimation.
Through interaction with the environment, the agent learns to emphasize high-SINR and geometrically informative beams while suppressing noisy or redundant ones, and the augmented WLS solver enforces geometric consistency during localization.
We also include supervised LSTM-based weighting (LSTM-WLS~\cite{Sbeity2023_LSTM}) as a reference that targets accuracy-optimal estimation in its supervised setting.
In contrast, the proposed DQN-WLS is designed for CSI-free and deployment-oriented operation with low runtime overhead, which is relevant to resource-constrained LEO payload scenarios.

\section{Conclusion}
This paper investigated a DQN-WLS framework for CSI-free multi-satellite positioning in LEO constellations that learns discrete satellite-weighting policies from pilot-level observations and geometric features.
By coupling a lightweight DQN policy with an augmented WLS estimator, the proposed method provides a deployment-oriented accuracy-runtime trade-off and preserves physics-consistent localization with clock-bias estimation.
Simulation results in representative settings demonstrate sub-meter accuracy with low computational overhead, suggesting that hybrid RL-WLS designs are a promising direction for CSI-free beam-based positioning in NTNs.
Extending the framework to broader beam counts, richer interference models, and more diverse geometries is a natural next step.

\end{document}